%% file: paper.tex
\documentclass[11pt]{article}
\usepackage{graphicx}

\newcommand{\BABARPubYear}    {02}

\newcommand{\BABARConfNumber} {027}
\newcommand{\SLACPubNumber} {9309}

\input{babarsym}

\input{symbols}

\input{definitions}

\long\def\inst#1{\par\nobreak\kern 4pt\nobreak
    {\it #1}\par\vskip 10pt plus 3pt minus 3pt}

\begin{document}
{\pagestyle{empty}

\begin{flushright}

\babar-CONF-\BABARPubYear/\BABARConfNumber \\
SLAC-PUB-\SLACPubNumber \\


July 2002 \\
\end{flushright}

\par\vskip 5cm

\begin{center}
\Large \bf A Search for {\Bknunu}
\end{center}
\bigskip

\begin{center}
\large The \babar\ Collaboration\\
\mbox{ }\\
July 24, 2002
\end{center}
\bigskip \bigskip

\begin{center}
\large \bf Abstract
\end{center}
\noindent
We present a preliminary search for the flavor-changing neutral current decay
{\ensuremath{\B^{+}\rightarrow K^{+}\nu\overline\nu\xspace}} using
$56$ million \FourS\ decays recorded with the
{\mbox{\slshape B\kern-0.1em{\smaller A}\kern-0.1em B\kern-0.1em{\smaller A\kern-0.2em R}}}
detector at the
SLAC $B$ factory. Identification of the
{\ensuremath{\B^{+}\rightarrow K^{+}\nu\overline\nu\xspace}} final state,
with two neutrinos, requires the
reconstruction of the companion $B$ in the event.  The companion $B$
is reconstructed in the decay mode
{\ensuremath{B^{-}\rightarrow D^{0}\ell^{-}\overline\nu~X}},
which provides both high efficiency and good purity.  The particles not
used in the reconstruction of
the companion $B$ are compared with the signature expected for
{\ensuremath{\B^{+}\rightarrow K^{+}\nu\overline\nu\xspace}} decays.
Two candidates are found in the data with an
expected background of 2.2 events.
Under the assumption that all candidates are signal events,
an upper limit on the branching fraction for
{\ensuremath{\B^{+}\rightarrow K^{+}\nu\overline\nu\xspace}}
of $9.4\,\times\,10^{-5}$ at 90\%\ confidence level is determined.

\vfill
\begin{center}
Contributed to the 31$^{st}$ International Conference on High Energy Physics,\\ 
7/24---7/31/2002, Amsterdam, The Netherlands
\end{center}

\vspace{1.0cm}
\begin{center}
{\em Stanford Linear Accelerator Center, Stanford University, 
Stanford, CA 94309} \\ \vspace{0.1cm}\hrule\vspace{0.1cm}
Work supported in part by Department of Energy contract DE-AC03-76SF00515.
\end{center}

\newpage
} 

\input authors_ICHEP2002.tex

\setcounter{footnote}{0}

\section{Introduction}
\label{sec:Introduction}

\input introduction.tex

\section{The dataset}
\label{sec:babar}

\input detector.tex

\section{Analysis method}
\label{sec:Analysis}

\input method.tex

\section{Physics results}
\label{sec:Physics}

\input physics-results.tex

\section{Acknowledgments}
\label{sec:Acknowledgments}

\input acknowledgements


\end{document}

%% file: babarsym.tex
\usepackage{relsize}
\def\babar{\mbox{\slshape B\kern-0.1em{\smaller A}\kern-0.1em
    B\kern-0.1em{\smaller A\kern-0.2em R}}}

\def\epem       {\ensuremath{e^+e^-}}

\def\taup       {\ensuremath{\tau^+}}

\def\tautau     {\ensuremath{\tau^+\tau^-}}
\def\ellm       {\ensuremath{\ell^-}}

\def\nunub {\ensuremath{\nu\overline \nu}}
\def\nub {\ensuremath{\overline \nu}}

\def\qqbar {\ensuremath{q\overline q}}

\def\uubar {\ensuremath{u\overline u}}

\def\ddbar {\ensuremath{d\overline d}}
\def\s  {\ensuremath{s}}

\def\ssbar {\ensuremath{s\overline s}}

\def\ccbar {\ensuremath{c\overline c}}

\def\piz   {\ensuremath{\pi^0}}

\def\pip   {\ensuremath{\pi^+}}
\def\pim   {\ensuremath{\pi^-}}

\def\kaon  {\ensuremath{K}}
\def\Kbar  {\ensuremath{\kern 0.2em\overline{\kern -0.2em K}}}

\def\Kp    {\ensuremath{K^+}}
\def\Km    {\ensuremath{K^-}}

\def\Kstar   {\ensuremath{K^*}}
\def\Kstarb   {\ensuremath{{\Kbar}\kern0.03em {\raise.18ex\hbox{$^*$}}}}

\def\Kzb   {\ensuremath{{\Kbar}\kern0.03em {\raise.1ex\hbox{$^0$}}}}
\def\KzKzb {\ensuremath{K^0 {\kern -0.16em \Kzb}}}
\def\Dz    {\ensuremath{D^0}}
\def\Dbar  {\ensuremath{\kern 0.2em\overline{\kern -0.2em D}}}
\def\Dzb   {\ensuremath{{\Dbar}\kern0.03em {\raise.3ex\hbox{$^0$}}}}
\def\DzDzb {\ensuremath{D^0 {\kern -0.16em \Dzb}}}

\def\Dstarb   {\ensuremath{{\Dbar}\kern0.03em {\raise.18ex\hbox{$^*$}}}}

\def\B     {\ensuremath{B}}
\def\Bbar  {\ensuremath{\kern 0.18em\overline{\kern -0.18em B}}}
\def\Bzyb   {\ensuremath{{\Bbar}\kern0.03em {\raise.3ex\hbox{$^0$}}}}
\def\Bzb   {\ensuremath{{\Bbar}\kern0.03em {\raise.3ex\hbox{$^0$}}}}
\def\Bu    {\ensuremath{B^+}}
\def\Bp    {\ensuremath{B^+}}
\def\Bub   {\ensuremath{B^-}}
\def\Bm    {\ensuremath{B^-}}

\def\BB    {\ensuremath{B\Bbar}} 
\def\BzBzb {\ensuremath{B^0 {\kern -0.16em \Bzb}}}

\mathchardef\Upsilon="7107
\def\Y#1S{\ensuremath{\Upsilon{(#1S)}}}

\def\FourS {\Y4S}

\mathchardef\Delta="7101
\mathchardef\Xi="7104
\mathchardef\Lambda="7103
\mathchardef\Sigma="7106
\mathchardef\Omega="710A
\def\Deltabar   {\ensuremath{\kern 0.25em\overline{\kern -0.25em \Delta}}{}}
\def\Lbar {\ensuremath{\kern 0.2em\overline{\kern -0.2em\Lambda\kern 0.05em}\kern-0.05em}{}}
\def\Sigbar{\ensuremath{\kern 0.2em\overline{\kern -0.2em \Sigma}}{}}
\def\Xibar{\ensuremath{\kern 0.2em\overline{\kern -0.2em \Xi}}{}}
\def\Obar{\ensuremath{\kern 0.2em\overline{\kern -0.2em \Omega}}{}}
\def\Nbar{\ensuremath{\kern 0.2em\overline{\kern -0.2em N}}{}}

\def\BR{{\ensuremath{\cal B}}}

\def\ev   {\ensuremath{\rm \,e\kern -0.08em V}}
\def\kev  {\ensuremath{\rm \,ke\kern -0.08em V}} 
\def\mev  {\ensuremath{\rm \,Me\kern -0.08em V}} 
\def\gev  {\ensuremath{\rm \,Ge\kern -0.08em V}} 
\def\gevc {\ensuremath{{\rm \,Ge\kern -0.08em V\!/}c}} 
\def\tev  {\ensuremath{\rm \,Te\kern -0.08em V}}
\def\mevc {\ensuremath{{\rm \,Me\kern -0.08em V\!/}c}} 
\def\gevcc{\ensuremath{{\rm \,Ge\kern -0.08em V\!/}c^2}} 
\def\mevcc{\ensuremath{{\rm \,Me\kern -0.08em V\!/}c^2}}

\def\invfb   {\ensuremath{\mbox{\,fb}^{-1}}}
\def\mus  {\ensuremath{\rm \,\mus}}

\def\mus        {\ensuremath{\,\mu{\rm s}}}    
\def\gsim{{~\raise.15em\hbox{$>$}\kern-.85em
          \lower.35em\hbox{$\sim$}~}}
\def\lsim{{~\raise.15em\hbox{$<$}\kern-.85em
          \lower.35em\hbox{$\sim$}~}}

\def\to                 {\ensuremath{\rightarrow}}

\def\pep2{PEP-II}

\def\chic1{\ensuremath{\chi_{c1}}}
\def\chic2{\ensuremath{\chi_{c2}}}
\def\chic3{\ensuremath{\chi_{c3}}}

\def\jetset74   {\mbox{\tt Jetset \hspace{-0.5em}7.\hspace{-0.2em}4}}

%% file: symbols.tex
\def\er #1 #2 { $#1 \pm #2$ }

\def\mpl #1 #2 #3 {Mod.~Phys.~Lett.~{\bf#1},\ #2 (#3)}
\def\npb  #1 #2 #3 {Nucl.~Phys.~B~{\bf#1},\ #2 (#3)}
\def\plb  #1 #2 #3 {Phys.~Lett.~B~{\bf#1},\ #2 (#3)}
\def\pr   #1 #2 #3 {Phys.~Rep.~{\bf#1},\ #2 (#3)}
\def\prd  #1 #2 #3 {Phys.~Rev.~D~{\bf#1},\ #2 (#3)}
\def\prl  #1 #2 #3 {Phys.~Rev.~Lett.~{\bf#1},\ #2 (#3)}
\def\RMP  #1 #2 #3 {Rev.~Mod.~Phys.~{\bf#1},\ #2 (#3)}
\def\zpc  #1 #2 #3 {Z.~Phys.~C~{\bf#1},\ #2 (#3)}
\def\nim  #1 #2 #3 {Nucl.~Instrum.~Methods~{\bf#1},\ #2 (#3)}
\def\nima  #1 #2 #3 {Nucl.~Instrum.~Methods~A.{\bf#1},\ #2 (#3)}
\def\epjc #1 #2 #3 {Euro.~Phys.~Jour~{\bf#1},\ #2 (#3)}
\def\rmp #1 #2 #3 {Rev.~Mod.~Phys~{\bf#1},\ #2 (#3)}
\def\npbps #1 #2 #3 {Nucl.~Phys.~B.~proc.~suppl~{\bf#1},\ #2 (#3)}
\def\progtp #1 #2 #3 {Prog.~Theo.~Phys~{\bf#1},\ #2 (#3)}

%% file: definitions.tex
\newcommand{\xspace}    {\hspace{0.1cm}}

\newcommand{\GeV}       {\ensuremath{\mathrm{GeV}}}

\newcommand{\bDlnu}     {\ensuremath{\Bm\to\Dz\ellm\nub~X}}

\newcommand{\bKpnunu}    {\ensuremath{\Bp \to \Kp\nunub\xspace}}

\newcommand{\bKznunu}    {\ensuremath{\Bzb \to \Kzb\nunub\xspace}}

\newcommand{\bKstarnunu}    {\ensuremath{\B \to \Kstar\nunub\xspace}}

\newcommand{\Dsl}{\ensuremath{D}\ell\xspace}

\newcommand{\btodlnu} {\ensuremath{B\to D\ell\nu X}}

\newcommand{\btodstzerolnu} {\ensuremath{B^{-}\to D^{*0}\ell^{-}\nub}}
\newcommand{\btoddstzerolnu} {\ensuremath{B^{-}\to D^{**0}\ell^{-}\nub}}

\newcommand{\btotaunu}  {\ensuremath{\Bu~\to~\taup\nu\xspace}}

\newcommand{\Bknunu}      {\ensuremath{\Bp\to\Kp\nunub\xspace}}

\newcommand{\bsnunu}      {\ensuremath{b\to\s\nunub\xspace}}

\newcommand{\dztokpi} {\ensuremath{D^{0}\rightarrow K^{-}\pi^{+}}}

%% file: authors_ICHEP2002.tex
\begin{center}
\small

The \babar\ Collaboration,
\bigskip

B.~Aubert,
D.~Boutigny,
J.-M.~Gaillard,
A.~Hicheur,
Y.~Karyotakis,
J.~P.~Lees,
P.~Robbe,
V.~Tisserand,
A.~Zghiche
\inst{Laboratoire de Physique des Particules, F-74941 Annecy-le-Vieux, France }
A.~Palano,
A.~Pompili
\inst{Universit\`a di Bari, Dipartimento di Fisica and INFN, I-70126 Bari, Italy }
% ...

\end{center}\newpage

%% file: introduction.tex
The investigation of flavor-changing neutral current (FCNC) decays is of
fundamental interest.  In the Standard Model (SM) these decays are 
forbidden at tree level, and occur only in loop diagrams.  As a result,
their rates are highly suppressed.  The SM
prediction for the FCNC decay {\bsnunu} is nearly free from strong
interaction effects and has very small theoretical
uncertainty.  An observation of this decay at a level significantly
above the SM prediction would provide unambiguous evidence for new physics.

Within the SM the decay {\bsnunu} proceeds through $W$ box
diagrams and $Z$ penguin diagrams.
The expected
branching fraction, summed over all neutrino species,
is~\cite{babar-physics-book}
\begin{equation}
{\BR}({\bsnunu}) = \left( 4.1 ^{+0.8}_{-1.0}\right) \times 10^{-5}\ \ .
\end{equation}
At present it does not appear to be feasible to search for the inclusive
decay {\bsnunu}; 
however, the decay {\bKpnunu} is tractable.\footnote{Charge conjugate 
modes are implied throughout.} 
The expected branching fraction for {\bKpnunu}, summed
over all neutrino species, is~\cite{btosnunu-theory2}
\begin{equation}
{\BR}({\bKpnunu}) = \left( 0.38 ^{+0.12}_{-0.06}\right) \times 10^{-5}\ \ .
\end{equation}
The best previous experimental limit is 
$ {\BR}({\bKpnunu}) < \, 2.4 \, \times\, 10^{-4}$ at 90\%\ confidence 
level~\cite{cleo-bknunu}.

%% file: detector.tex
The data used in this analysis were collected with the \babar\ detector,
which is described elsewhere~\cite{ref:babar}, at the \pep2\ storage ring.
The integrated luminosity used in this analysis is $50.7\invfb$ recorded at the
\FourS\ resonance, corresponding to $56.3\,\times\,10^6$ $\BB$ events,
and $6.4\invfb$ taken at energies just below $\BB$ threshold.
Simulated data samples for the processes
$\epem\to\BB$, $\epem\to\qqbar$ ($q=u,\ d,\ s\ {\rm or}\ c$) and
$\epem\to\tautau$, in quantities comparable to the data, are used
to study backgrounds.  A sample of $280\,000$ simulated $\Bu\Bub$ events with
$\bKpnunu$ and the other $B$ decaying generically have also been analyzed.
The simulation of $\bKpnunu$ decay is based on the form factor model
in Ref.~\cite{btosnunu-theory2}.

%% file: method.tex
The presence of two neutrinos in the final state makes the search
for {\bKpnunu} difficult, since no kinematic constraints can be applied 
to the signal $B$.  The strategy adopted in this analysis is
to reconstruct exclusively the decay of one of the $B$ mesons in the
event, referred to as the ``tag'' $B$, and to compare the remaining
particle(s) in the event with the signature expected for the decay {\bKpnunu}.
The low multiplicity of the signal decay greatly reduces
the combinatorial background in the tag reconstruction, allowing the
use of decay modes that would not be sufficiently clean in other
circumstances.  These considerations lead to the use of the
semileptonic decay {\bDlnu} for the reconstruction of the tag $B$.
The $X$ system is kinematically constrained  to be either nothing or 
a low-momentum pion or photon from a higher mass charm state. 
The {\Dz} is reconstructed in the {\Km\pip}, {\Km\pip\pim\pip}
and {\Km\pip\piz} modes.  This method results in roughly 0.5\%\ of
{\Bm} decays being reconstructed as tags.  Note that particles from
the tag $B$ that escape detection will not affect the sensitivity of the 
analysis to {\Bknunu} events; 
the reconstructed $\Dsl$ needs to be a correct, but not
complete, subset of the particles produced in the tag $B$ decay.  The
feed-down from higher-mass charm states often results in good tags in
this sense, and thus in an enhanced tagging efficiency.

The event selection proceeds as follows.  Selected hadronic events are 
required to have an identified electron or muon with a 
momentum above $1.3$ GeV/c in the \FourS\ rest frame.  The electron
identification is based on quantities from the electromagnetic
calorimeter (EMC), the ring-imaging Cherenkov detector (DIRC)
and the gas (DCH) and silicon (SVT) tracking devices.
The muon identification uses information from the instrumented
flux return (IFR) in addition to the devices listed previously.
\begin{figure}[htbp]
\hspace{-0.25\textwidth}\makebox[1.5\textwidth][c]{
  \resizebox{0.33\textwidth}{!}
{\includegraphics{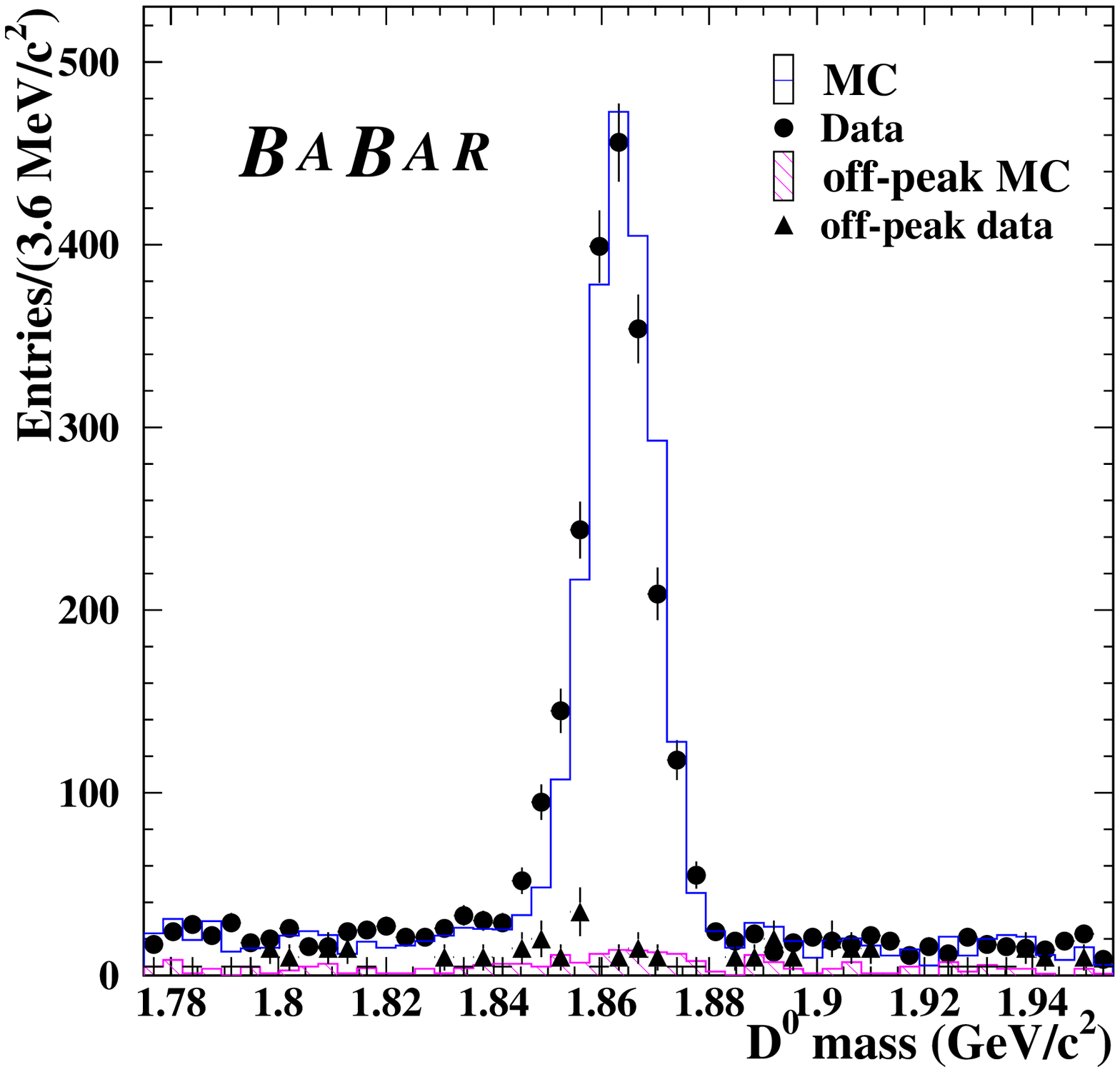}}
  \resizebox{0.33\textwidth}{!}
{\includegraphics{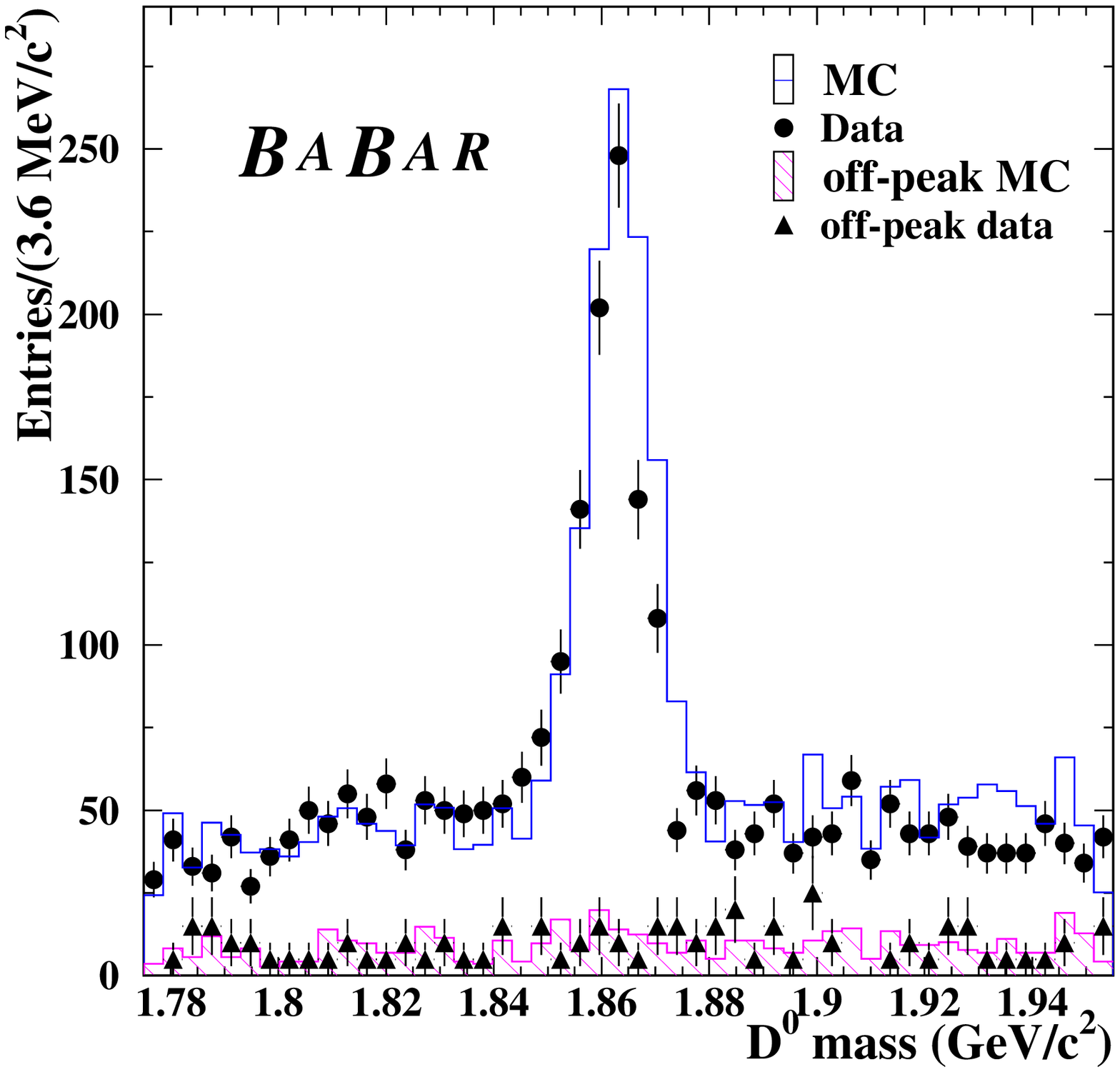}}
  \resizebox{0.33\textwidth}{!}
{\includegraphics{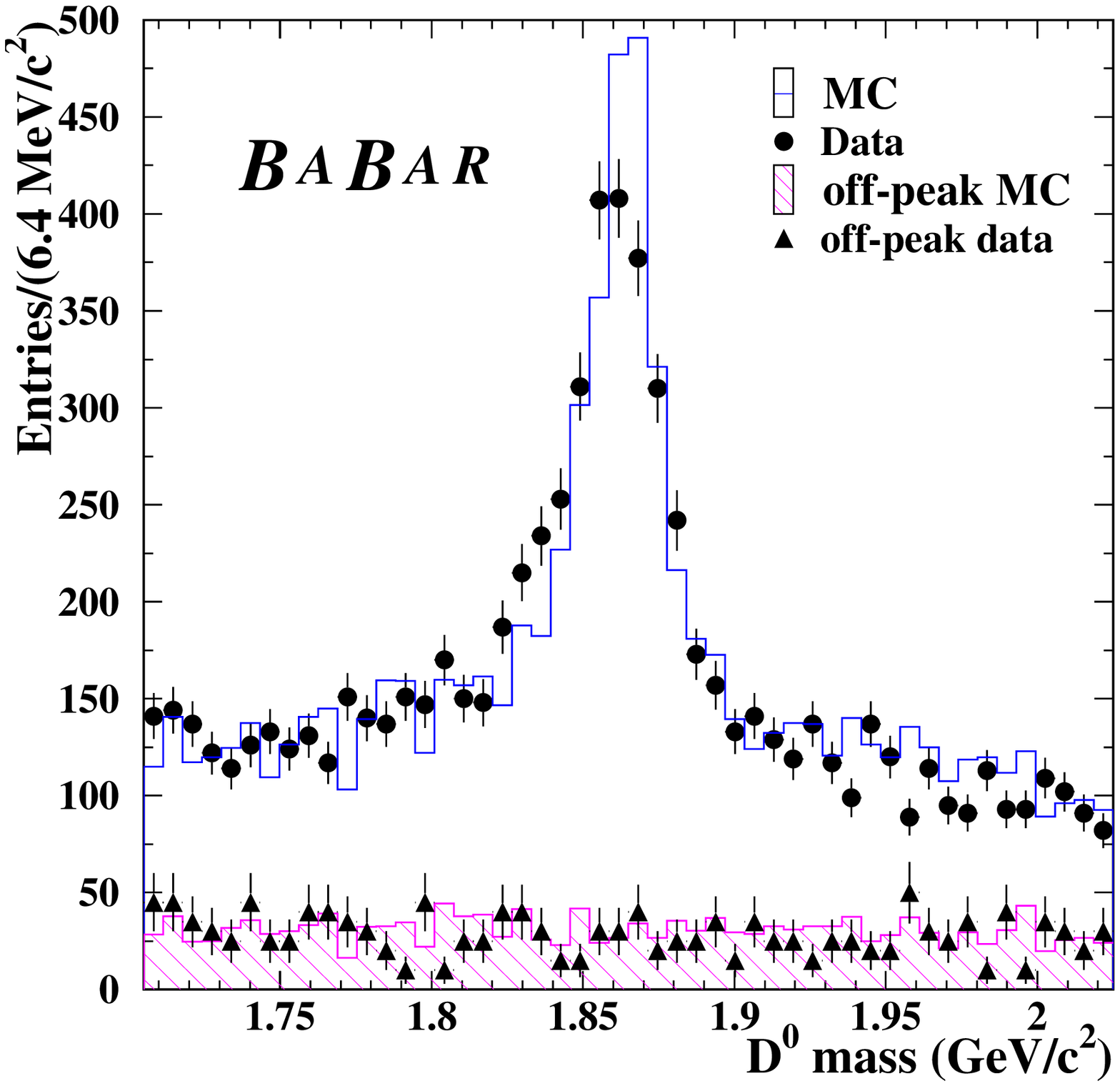}}
}
\begin{center}
 \caption{The candidate {\Dz} invariant mass distributions are shown, from left to 
right, in the {\Km\pip},
{\Km\pip\pim\pip} and {\Km\pip\piz} modes for data (points)
 and simulation (histogram), for 
events with no more than
three charged tracks and less than $1\,\GeV$ of neutral energy
not assigned to the tag $B$ candidate.
Events are required to have no more than three charged tracks not associated 
with the tag $B$ in order to mimic the low multiplicity of the signal 
while maintaining adequate statistics in the plots.
The off-resonance distributions have been scaled to the on-peak data 
luminosity.
 \label{fig:d0mass}}
\end{center}
\end{figure}
Loose consistency requirements are placed
on the charged particle vertices for the $\Dz$ and
$\Dz\ellm$ candidates.  The following kinematic requirements are imposed:
$p^*_{\Dz}>0.5\,\GeV/c$, $m_{\Dz\ellm}>3\,\GeV/c^2$ and
$-2.5<\cos\theta_{B,\Dsl}<1.1$, where $p^*_{\Dz}$ is the momentum of the
{\Dz} in the \FourS\ frame, $m_{\Dz\ellm}$ is the mass of the 
{\Dz\ellm} combination and
\begin{equation}
\cos\theta_{B,\Dsl} = \frac{2\,E_{{\B}} E_{{\Dsl}}
 -m^2_{{\B}} - m^2_{{\Dsl}}}
{2\,|\vec{p}_{{\B}}| |\vec{p}_{{\Dsl}}|   }\ \ .
\end{equation}
Here $E_{\B}$ and $|\vec{p}_{\B}|$ are respectively
the energy of and magnitude of
the momentum of the $\B$ meson in the \FourS\ frame.
$E_{\B}$ is one half of the center-of-mass energy of the 
$e^{+}e^{-}$ initial state, and $|\vec{p}_{\B}|$ is 
$\sqrt{(E_{\B}^{2}-m_{\B}^{2})}$.
The upper limit on $\cos\theta_{B,\Dsl}$ is 1.1 to account for
resolution on the measurement (the signal cannot exceed 1).
The lower limit is relaxed to increase efficiency for the feed-down 
from decays of the type {\btodstzerolnu} and {\btoddstzerolnu}.
The requirement on $\cos\theta_{B,\Dsl}$ is
the most important for restricting the
kinematics of the $\Dz\ellm$ to be consistent with coming from
a semileptonic $B$ decay.   In cases where more than one $\Dz\ellm$
candidate is reconstructed, the one with the smallest value of
$|\cos\theta_{B,\Dsl}|$ is used.  The reconstructed $\Dz$ invariant
mass distributions 
are shown in Fig.~\ref{fig:d0mass}.

\begin{figure}[htbp]
\hspace{-0.25\textwidth}\makebox[1.5\textwidth][c]{
  \resizebox{0.465\textwidth}{!}
{\includegraphics{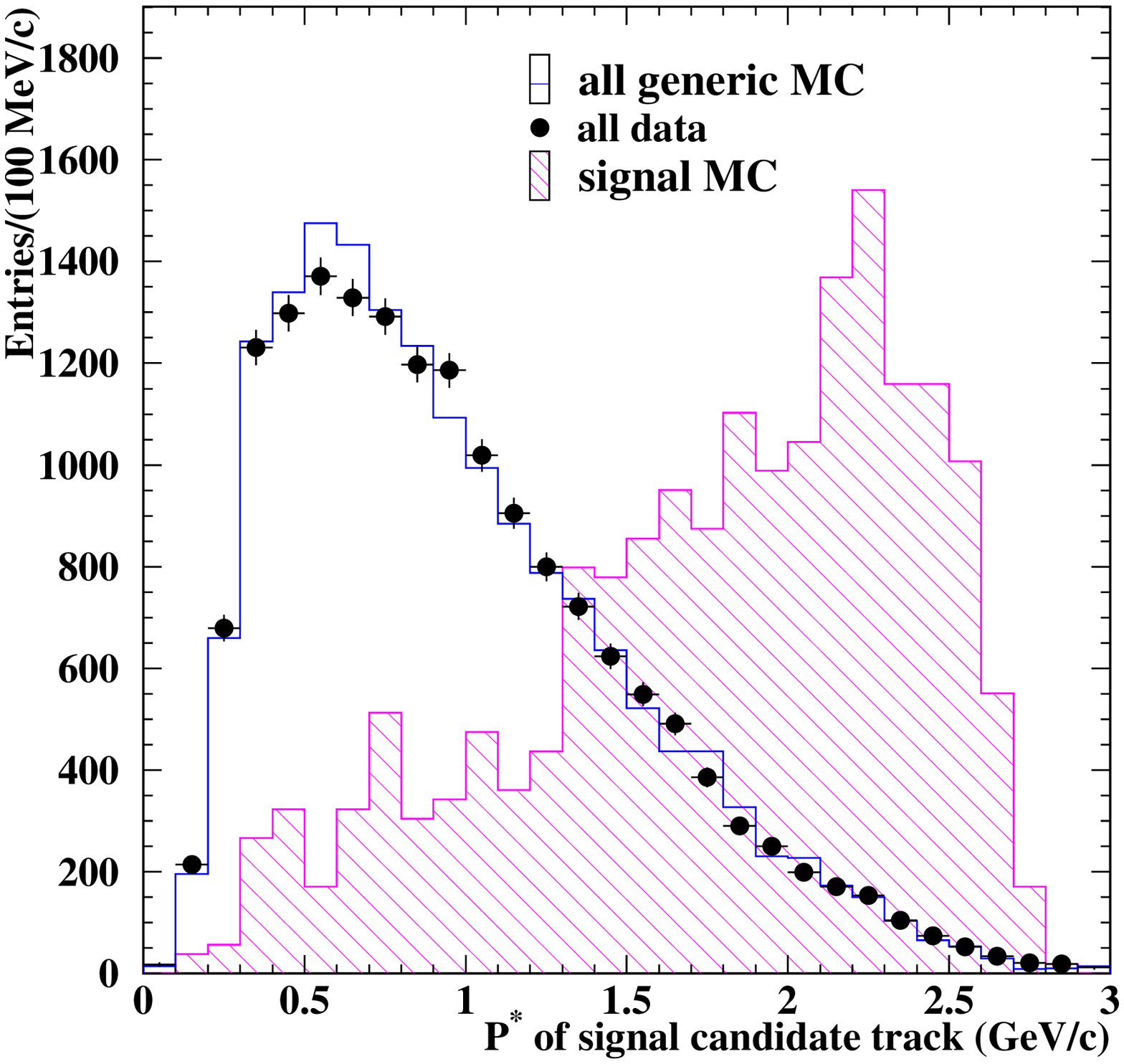}}\hspace{1cm}
\resizebox{0.465\textwidth}{!}
{\includegraphics{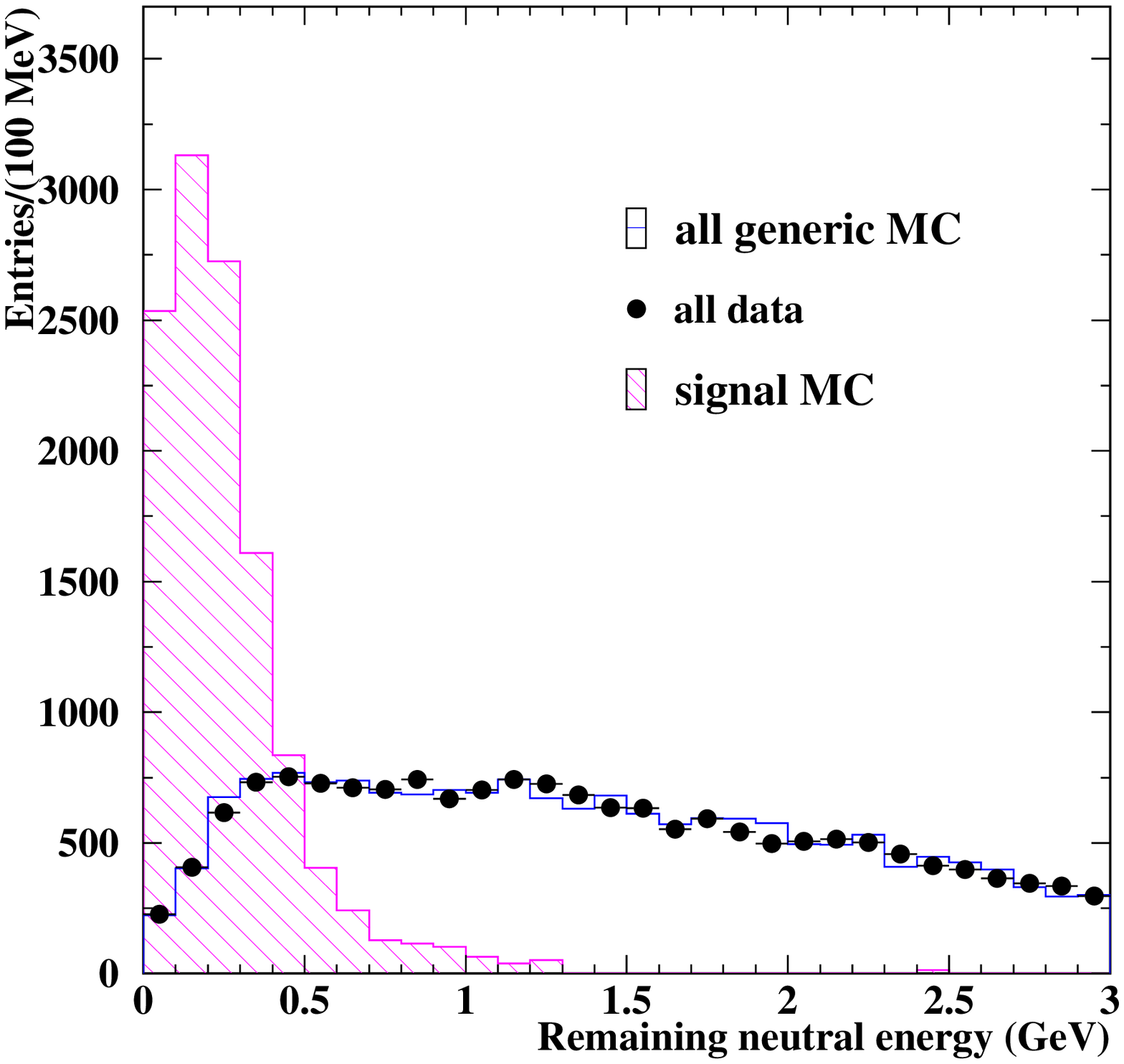}}
}
\begin{center}
 \caption{The distributions of $p^*_K$ and $E_{\rm left}$ for
simulated signal and background events.
Events with no more than
three charged tracks and less than $1\,\GeV$ of neutral energy
not assigned to the tag $B$ candidate are used for the plot on the left
whereas the neutral energy requirement is relaxed to be less than $3\,\GeV$
for the plot on the right.
The generic MC distribution has been scaled to the on-peak data 
luminosity with an arbitrary scale factor applied to the signal MC
distribution.
\label{fig:signal-selection}}
\end{center}
\end{figure}

Once the tag $B$ is selected, additional requirements are placed on 
the remaining particles in
the event.  There must be exactly one charged track in the event that
is not part of the tag $B$, its charge must be opposite to that
of the tag lepton, and it must satisfy
the particle identification criteria for charged kaons,
which are based on information
from the DIRC and tracking system.  
The momentum spectrum, in the \FourS\ rest frame, for the kaon
from {\Bknunu} decays peaks near the upper kinematic limit while
the spectrum for background peaks at low momentum; the signal
kaon candidate is thus required to
satisfy $p_K^*>1.5\,\GeV/c$ (see Fig.~\ref{fig:signal-selection}).
The angle $\theta^{*}_{K,\ell}$ between the charged lepton
and the signal kaon is isotropically distributed in signal events, since
these particles originate from different $B$ mesons, while the background
from $\epem\to\qqbar$ and $\epem\to\tautau$ peaks forward and backward
in this angle; we require $-0.9<\cos\theta^{*}_{K,\ell}<0.8$.  
In addition to the above requirements on charged tracks, we use information
from the EMC and IFR to limit additional neutral particles in the event.
The {\bKpnunu} signal
leaves very little neutral energy in the detector and does not
contain any neutral hadrons.  We therefore require that the number
of IFR clusters consistent with neutral hadrons ($N_{\rm IFR}$) be zero,
and that the energy deposited in the EMC, once the daughters from
the $\Dsl$ have been removed, (referred to as $E_{\rm left}$ or 
remaining neutral energy)
satisfies $E_{\rm left}<0.5\,\GeV$
(see Fig.~\ref{fig:signal-selection}).

\begin{figure}
\begin{center}
\parbox{3.1in}
{\includegraphics[width=3.0in,height=3.0in]{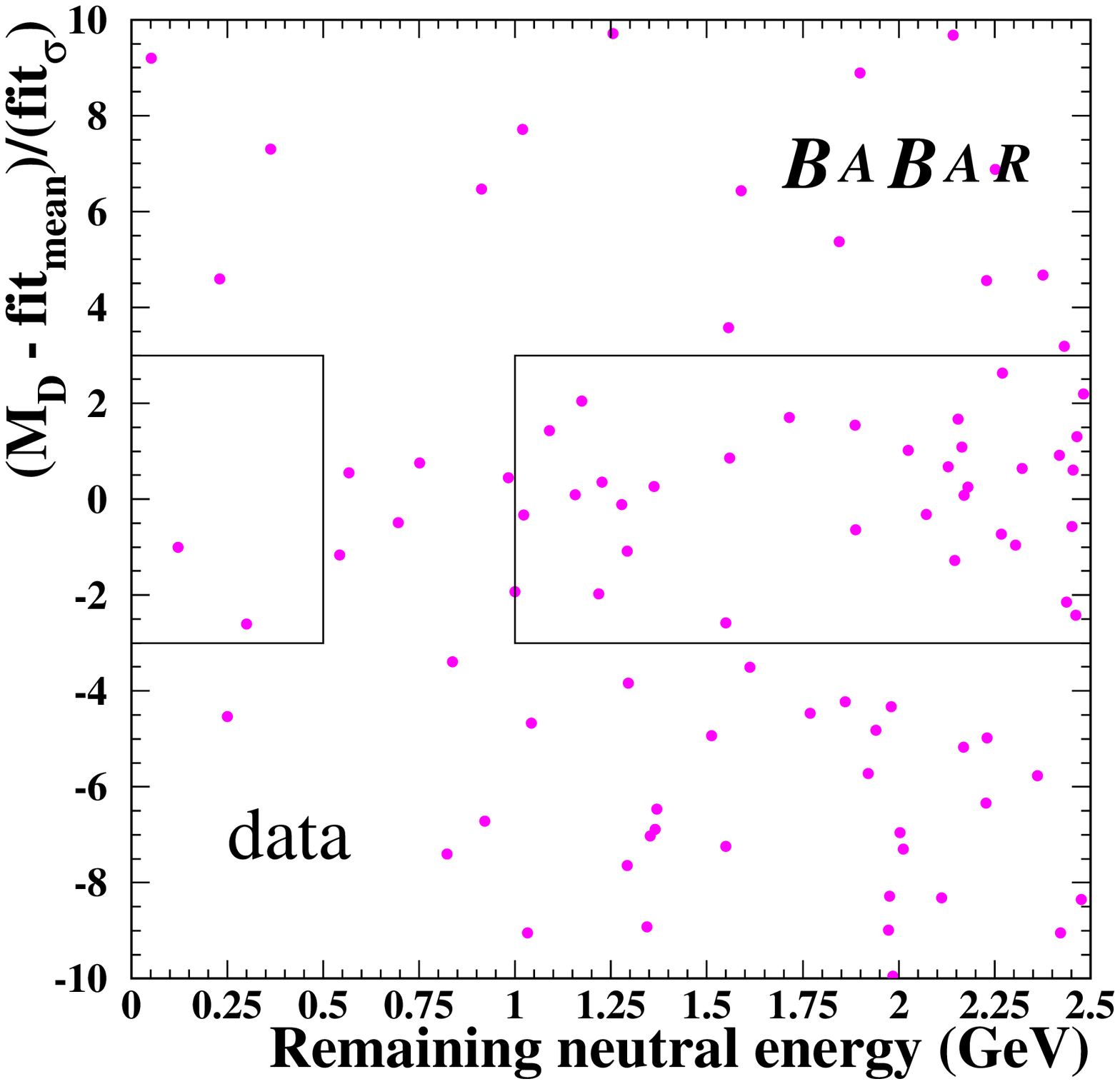}}
\parbox{3.1in}
{\includegraphics[width=3.0in,height=3.0in]{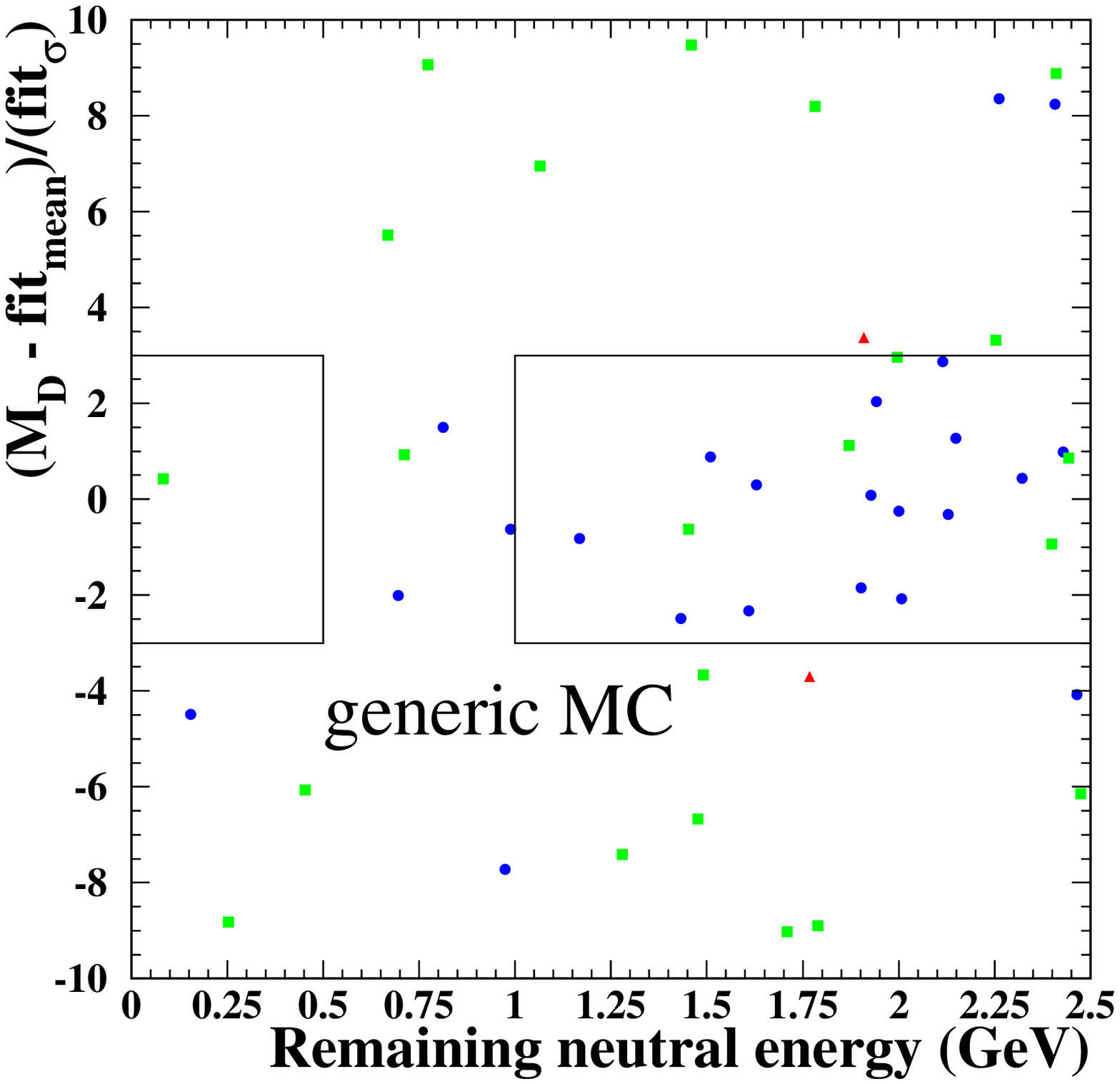}}
\end{center}
\begin{center}
\parbox{3.3in}
{\includegraphics[width=3.0in,height=3.0in]{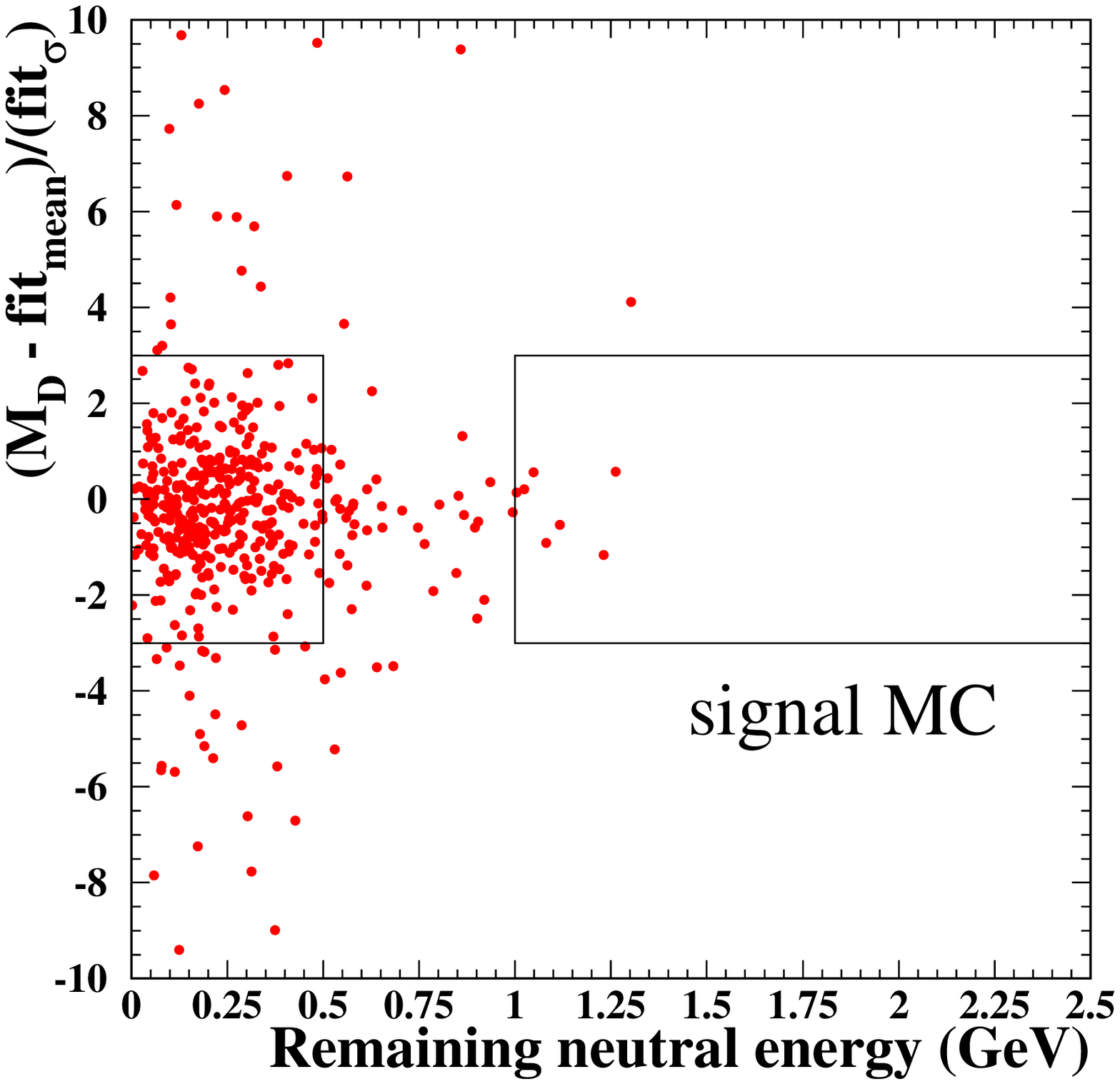}}
\end{center}
\begin{center}
 \caption{The distribution of events in the
 $(m_D - m_D^{\rm fit})/\sigma_D^{\rm fit}$ versus $E_{\rm left}$
 plane for on-peak data, generic $\BB$ and continuum Monte
 Carlo and signal Monte Carlo.
 In the generic Monte Carlo plot the circles show the contribution
 from $\BB$ events, the squares show the contribution from $\ccbar$
 and the triangles show the contribution from $\uubar/\ddbar/\ssbar$.
 The MC has not been scaled to the data luminosity.
 \label{fig:signalplane}}
\end{center}
\end{figure}

\begin{table}
\begin{center}
\caption{The number of events passing the selection criteria for on-peak data, 
on-peak Monte Carlo contributions, off-peak data, off-peak Monte Carlo 
contributions and \bKpnunu\ signal Monte Carlo efficiency.
The number of events in the Monte Carlo sample are scaled to the equivalent 
luminosity in data.
The values include the correction factors for tag efficiency, $E_{\rm left}$
and  $N_{\rm IFR}$ referred to in the text.
\label{tab:eff}}
\vspace{0.3cm}
\begin{tabular}{| l || r | r || r | r || r |}
\hline
\multicolumn{1}{|c||}{ } & \multicolumn{2}{|c||}{On-peak $(50.7\invfb)$}   & \multicolumn{2}{|c||}{Off-peak $(6.4\invfb)$} & \multicolumn{1}{|c|}{signal MC}    \\
\hline
 Requirement & {data yield}  & {MC yield}
             & {data yield}  & {MC yield} & effic $\cdot 10^4$ \\
\hline
 Tag, no extra tracks         & 8998 & 8525.7 & 415 & 389.9 & 34.3 \\
 Kaon identification          &  717 &  707.4 &  49 &  46.8 & 24.3 \\
 $\cos\theta^{*}_{K,\ell}$    &  485 &  486.2 &  32 &  25.0 & 20.9 \\
 $p^*_K$                      &  101 &   89.4 &   7 &   5.1 & 14.2 \\
 $N_{\rm IFR}$                &   79 &   72.5 &   6 &   4.4 & 12.0 \\
 \hline
 $E_{\rm left}$ sideband      &   34 &   27.4 &   3 &   1.4 &  0.2 \\
 \hline
{\Dz} mass sideband           &    4 &    7.1 &   1 &   0.8 &  2.0 \\
 \hline
 Signal box                   &    2 &    2.2 &   0 &   0.3 & 10.3 \\
\hline
\end{tabular}
\end{center}
\end{table}
\vspace{-0.4cm}
The yields in the signal and sideband regions at each stage in the application of the selection criteria
are given in Table~\ref{tab:eff}
for the on-peak data and background Monte Carlo, along with the
efficiency for the signal Monte Carlo.
The distribution of events in the search plane defined by the 
variables\footnote{The quantities 
$m_D^{\rm fit}$ and $\sigma_D^{\rm fit}$
are the mean and sigma from Gaussian fits to the $\Dz$ invariant mass
spectrum.  Separate values are calculated for each $\Dz$ decay mode in
data and simulation.}
$E_{\rm left}$ and $(m_D - m_D^{\rm fit})/\sigma_D^{\rm fit}$ is
shown in Fig.~\ref{fig:signalplane}.  
The signal box is defined by the requirements
$E_{\rm left}<0.5\,\GeV$ and $|m_D^{} - m_D^{\rm fit}|<3\sigma_D^{\rm fit}$.
The expected background from the Monte Carlo is 2.2 events.

In order to minimize experimental bias, the signal region was hidden until the 
selection criteria were finalized.
In order to evaluate how well the simulation describes
the data, we define auxiliary samples.  Two sideband regions
are studied: the $\Dz$ mass sideband, defined by the conditions
$|m_D - m_D^{\rm fit}|>3\sigma_D^{\rm fit}$ and $E_{\rm left}<0.5\,\GeV$, 
and a sideband where the additional neutral energy is required
to be in the range $1.0<E_{\rm left}<2.5\,\GeV$.  
The $\Dz$ mass sideband contains incorrectly reconstructed $B$~decays
and continuum events whereas the $E_{\rm left}$ sideband is 
sensitive to correctly reconstructed $B$~tags 
where the other~$B$~leaves only a single detected charged track and
substantial missing energy, often in the form of neutral hadrons.
The event yields in these regions are also listed in Table~\ref{tab:eff}.

In addition to the sideband samples, we use ``double-tagged'' events,
in which both {\Bu} and {\Bm} mesons are reconstructed as {\btodlnu}, to
quantify the uncertainty in the efficiency of several of our signal
criteria.  We reconstruct double-tagged events by finding a
suitable $\Dz\ellm$ candidate where the $\Dz$ decays to $\Km\pip$,
and then looking for a second $\Dzb\ell^+$ candidate in any of the
accepted $\Dzb$ modes.  No particle is assigned to more than
one of the $D\ell$ candidates.  In addition it is required that
the event contain
no charged tracks that are not assigned to a $D\ell$ candidate.

\begin{figure}[htbp]
\hspace{-0.25\textwidth}\makebox[1.5\textwidth][c]{
  \resizebox{0.472\textwidth}{!}
{\includegraphics{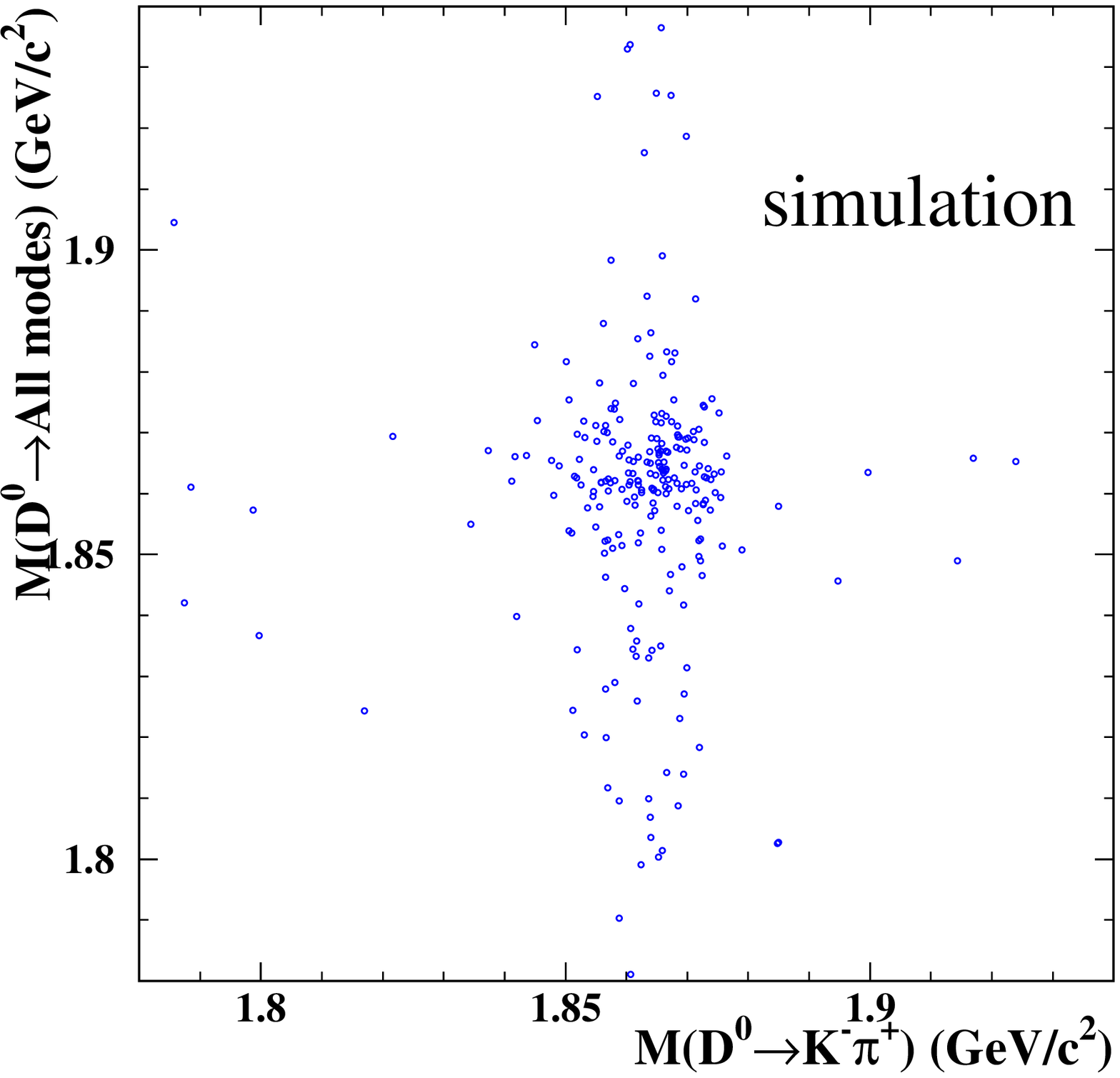}}
  \resizebox{0.472\textwidth}{!}
{\includegraphics{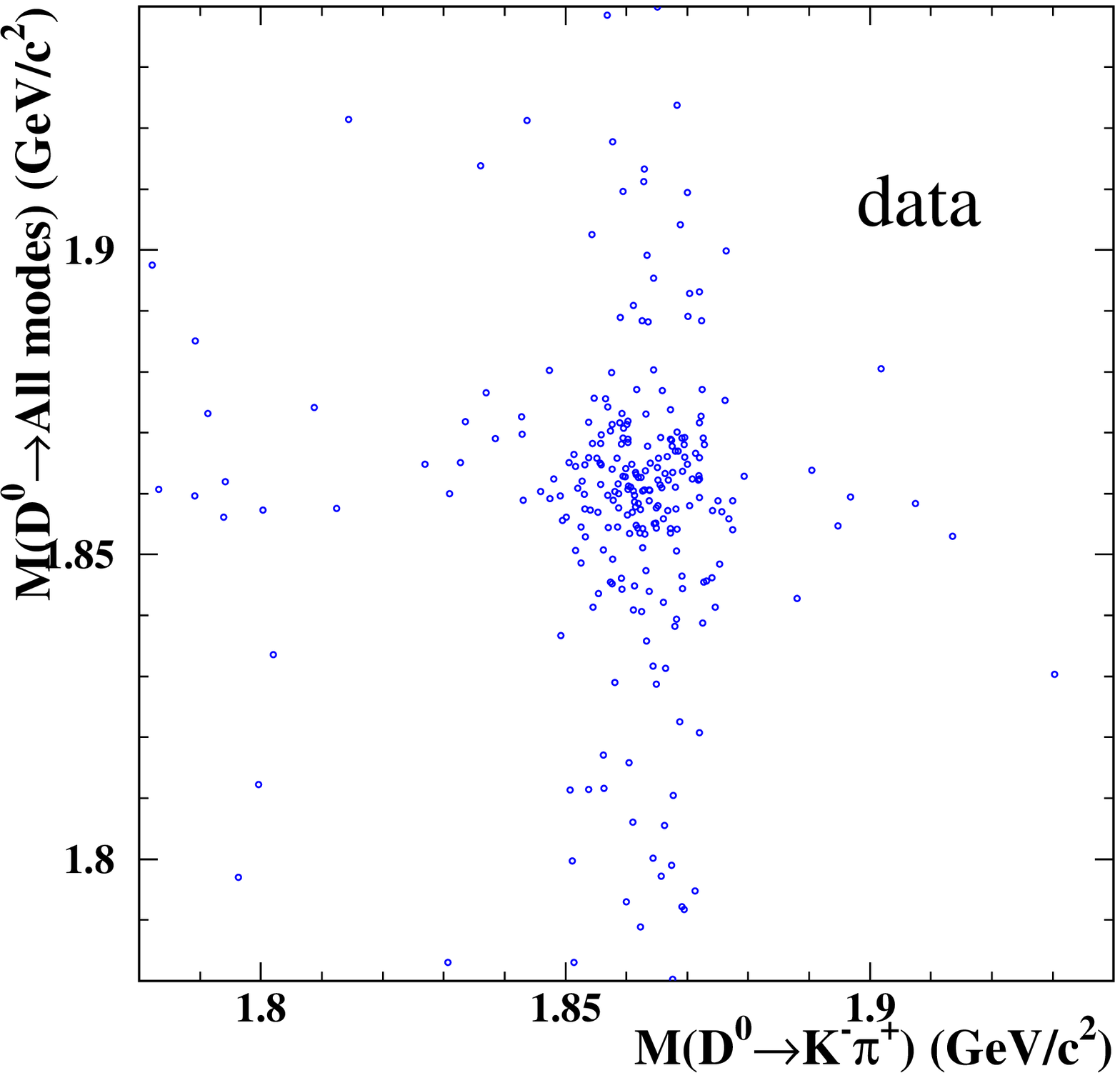}}
}
\begin{center}
 \caption{
Mass of the {\Dzb} candidate decaying to all three modes considered
versus the {\dztokpi} mass for events in which both $B$ mesons are reconstructed
 in the $\Dsl\nu X$ decay mode and no additional charged particles
 are recorded.  The plot on the left (right) shows the results
 from the simulation (on-peak data).
 \label{fig:dtag-mass}}
\end{center}
\end{figure}

The reconstructed invariant masses of the {\Dz} and $\Dzb$ candidates
in double-tagged events that satisfy the above criteria
are shown in Fig.~\ref{fig:dtag-mass} for data and for Monte Carlo.  
After subtracting
combinatorial background, the number of double-tagged events satisfying the
requirement that $|m_D - m_D^{\rm fit}| < 3\sigma_D^{\rm fit}$ for each
$D$ candidate
is $148\pm 15$ in data and $175\pm 16$ in the Monte Carlo 
sample.\footnote{The number of events in the Monte Carlo sample has been 
scaled to the on-peak data luminosity.}
The number of double-tagged events per $\invfb$ in the data is $0.85\pm 0.11$
times the rate in the simulation.  This factor is roughly the square
of the data/Monte Carlo efficiency ratio for
the tag efficiency (including the requirement
that there be no additional charged tracks associated with the tag - see
the first entry in Table~\ref{tab:eff}).
The signal efficiency is therefore corrected by a factor
$0.92\pm 0.06$, where the uncertainty is taken as a systematic error.

\begin{figure}[htbp]
\hspace{-0.25\textwidth}\makebox[1.5\textwidth][c]{
  \resizebox{0.472\textwidth}{!}
{\includegraphics{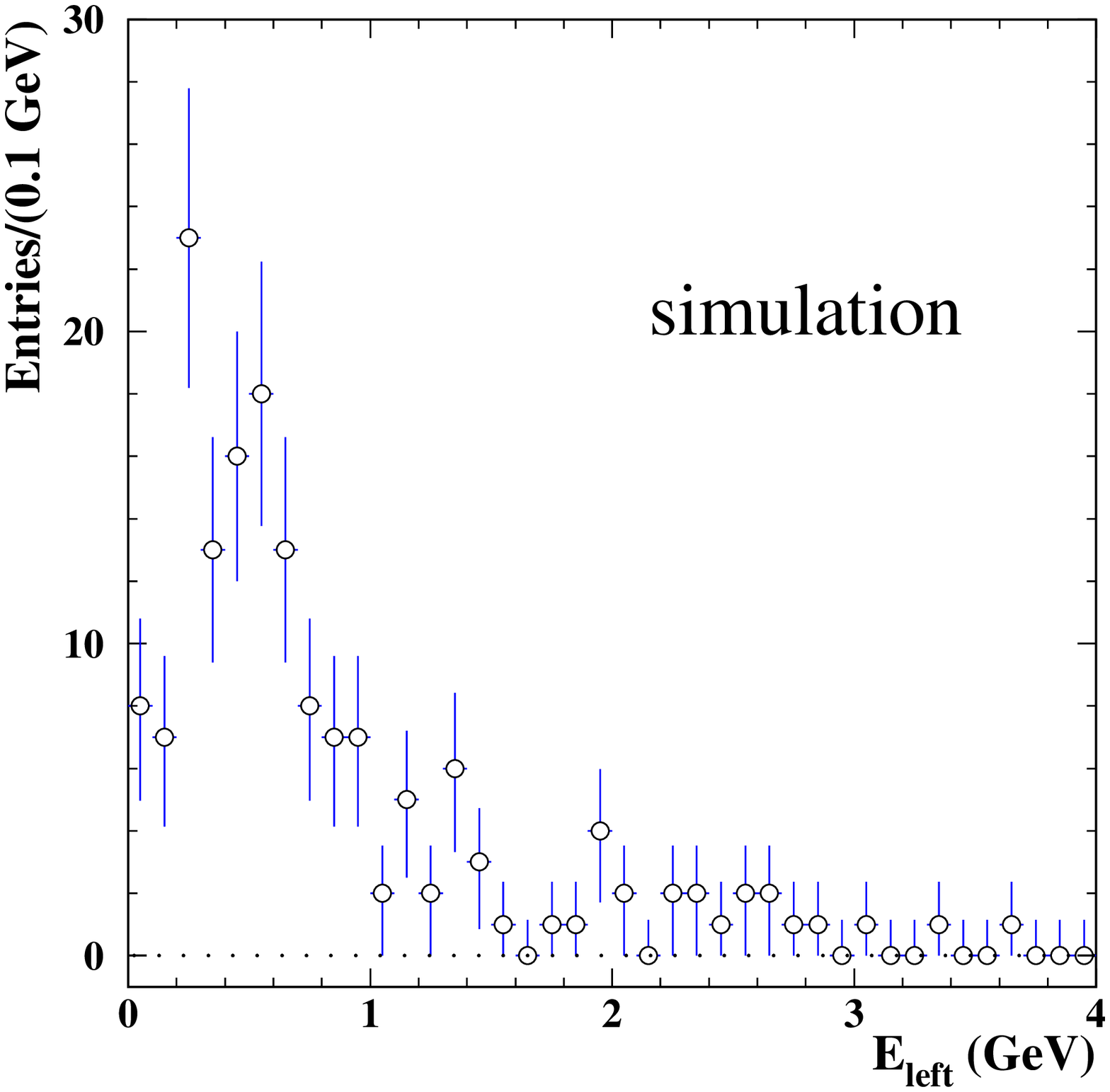}}
  \resizebox{0.472\textwidth}{!}
{\includegraphics{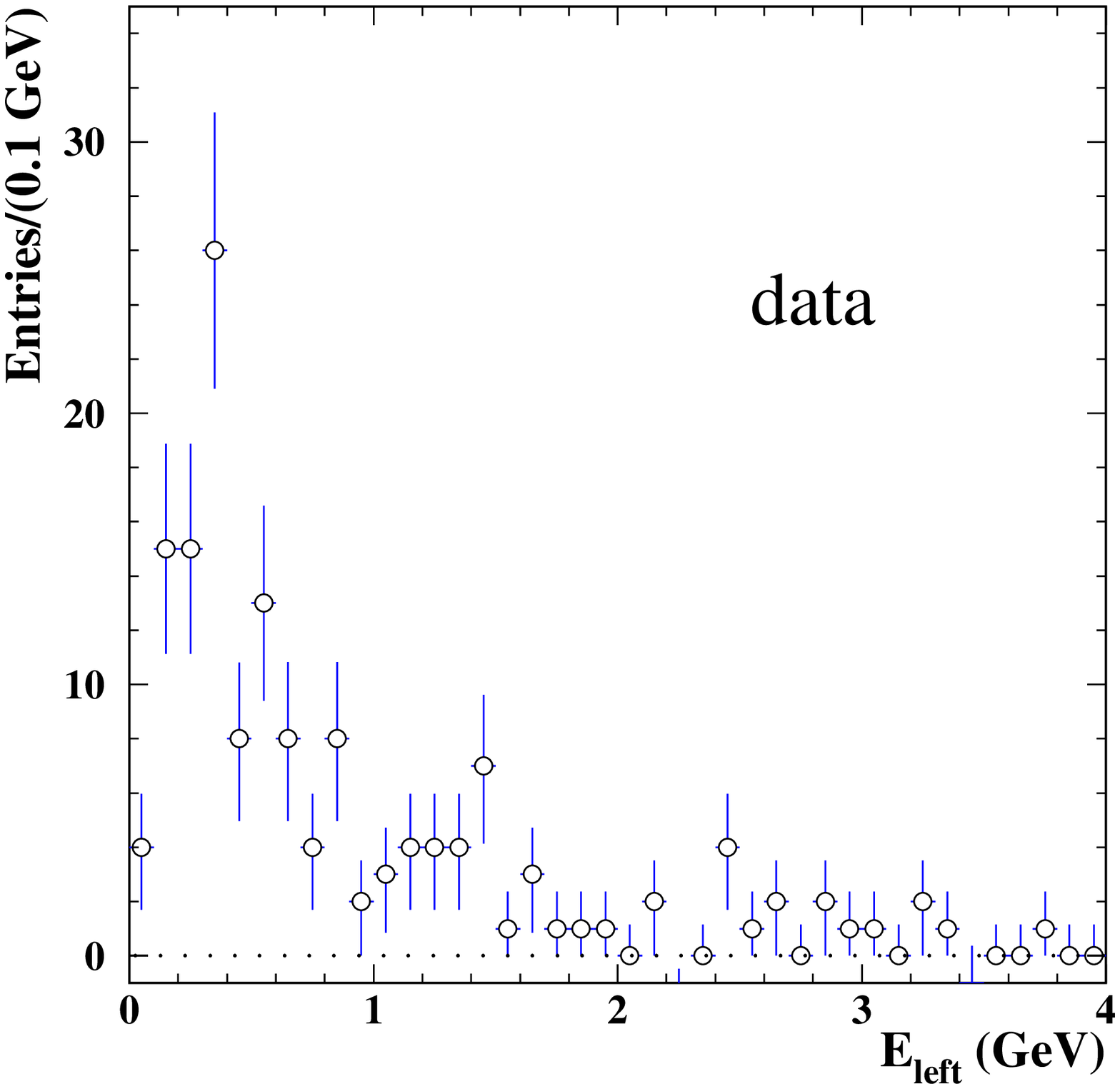}}
}
\begin{center}
 \caption{The distribution of $E_{\rm left}$ for ``double-tagged''
 events where both $B$ mesons are reconstructed
 in the $\Dsl X\xspace$ decay mode and no additional charged particles
 are recorded.  The plots on the left (right) show the
 distribution from simulation (on-peak data).
 \label{fig:dtag-eleft}}
\end{center}
\end{figure}

The double-tagged events also allow a study of how well the variables
$N_{\rm IFR}$ and $E_{\rm left}$ are simulated.  Figure~\ref{fig:dtag-eleft}
shows the distribution of the $E_{\rm left}$ variable in 
the double-tagged events;
the $\Dz$ mass sidebands have been used to subtract the combinatorial
background.  The mean values of $E_{\rm left}$ in the data and simulation
are $0.91\pm 0.08\,\GeV$ and $0.84\pm 0.07\,\GeV$, respectively.
The fraction of double-tagged events
satisfying the requirement $N_{\rm IFR}=0$ is $0.87\pm 0.03$ in data
and $0.93\pm 0.02$ in simulation. These comparisons are used to
adjust the simulated signal efficiencies and assign systematic errors.

\begin{table}
\begin{center}
\caption{A summary of the systematic uncertainties on {\BR}({\Bknunu}).
$\delta\epsilon/\epsilon$ is the relative uncertainty on the
overall efficiency.}
\begin{tabular}{| c | c |}
\hline
Quantity    &  $\delta\epsilon/\epsilon$[$\%$]  \\
\hline
{\BB}-counting              &  1.1   \\
Tagging efficiency          &  6.0   \\
{\kaon} selection           &  2.0   \\
$\cos\theta^{*}_{K,\ell}$   &  --    \\
$E_{\rm left}$              &  4.3   \\
$N_{\rm IFR}$               &  3.6   \\
{\kaon} momentum            &  1.8   \\
\hline
\end{tabular}
\label{tab:systematics}
\end{center}
\end{table}

Systematic uncertainties on the efficiency of selection criteria based on 
the total number of events with \FourS\ mesons,
tagging efficiency, {\kaon} selection and momentum, $E_{\rm left}$ and
$N_{\rm IFR}$ have all been studied.
The total relative uncertainty on the selection efficiency is found to be
$\delta\epsilon/\epsilon = 8.7\%$ where the tagging efficiency and
$E_{\rm left}$ contribute the largest uncertainties. The systematic 
uncertainties are summarised in Table~\ref{tab:systematics}.

%% file: physics-results.tex
The signal region was unblinded to reveal two events, consistent with 
the 2.2 events predicted with the simulation. 
The number of $\bKpnunu$ candidates in the data is thus compatible 
with the background expectation.  For the purpose of setting an upper
limit, each candidate is assumed to be signal.  The Poisson upper limit
for 2 events is 5.3.  This upper limit must be modified to account for
the uncertainty in the efficiency.  Using the prescription advocated
in~\cite{statisticsAWG} increases the upper limit to 5.4 events,
from which we find
\begin{equation}
{\BR}({\bKpnunu}) < 9.4 \times 10^{-5} \xspace\ \xspace\ \mbox{(preliminary)}
\end{equation}
at 90\%\ confidence level.

The background at present appears to be mostly combinatorial, based for example on 
the lack of any $\Dz$ peak in the continuum in Fig.~\ref{fig:d0mass}.
Further refinements may enable this background to be suppressed;
the combinatorial component of the background can also be subtracted
in the future.

The approach used in this analysis can easily be extended to
$\bKznunu$ and $\bKstarnunu$ as well as to $\btotaunu$.

%% file: acknowledgements.tex
We are grateful for the 
extraordinary contributions of our \pep2\ colleagues in
achieving the excellent luminosity and machine conditions
that have made this work possible.
The success of this project also relies critically on the 
expertise and dedication of the computing organizations that 
support \babar.
The collaborating institutions wish to thank 
SLAC for its support and the kind hospitality extended to them. 
This work is supported by the
US Department of Energy
and National Science Foundation, the
Natural Sciences and Engineering Research Council (Canada),
Institute of High Energy Physics (China), the
Commissariat \`a l'Energie Atomique and
Institut National de Physique Nucl\'eaire et de Physique des Particules
(France), the
Bundesministerium f\"ur Bildung und Forschung
(Germany), the
Istituto Nazionale di Fisica Nucleare (Italy),
the Research Council of Norway, the
Ministry of Science and Technology of the Russian Federation, and the
Particle Physics and Astronomy Research Council (United Kingdom). 
Individuals have received support from 
the A. P. Sloan Foundation, 
the Research Corporation,
and the Alexander von Humboldt Foundation.